\documentclass[apjl]{emulateapj}
\usepackage{apjfonts,natbib,epsfig}
\input epsf

\usepackage{wrapfig}

\newcommand{\beq}{\begin{equation}}
\newcommand{\beqa}{\begin{eqnarray}}
\newcommand{\eeq}{\end{equation}}
\newcommand{\eeqa}{\end{eqnarray}}

\newcommand{\lsim}{\lesssim}
\newcommand{\gsim}{\gtrsim}

\newcommand{\vect}[1]{\mbox{\boldmath${#1}$}}
\newcommand{\lmk}{\left(}
\newcommand{\rmk}{\right)}

\newcommand{\cm}{{\cal M}}

\begin{document}
\title{Detecting Planets around Compact Binaries with Gravitational Wave Detectors
in  Space}  
\author{Naoki Seto}
\affil{National Astronomical Observatory, 2-21-1
Osawa, Mitaka, Tokyo, 181-8588, Japan}

\begin{abstract}

I propose a method to detect planets around compact binaries that are
 strong sources of gravitational radiation. This approach is to measure
 gravitational-wave phase modulations induced by the planets, and its prospect
 is studied with a Fisher matrix analysis.
 I find that, using the Laser
Interferometer Space Antenna (LISA),     planets can be searched for around
 $\sim 3000$ Galactic double white dwarfs with detection limit $\gsim 4 M_J$ ($M_J\sim  2\times 10^{30}$g: the Jupiter mass). With  its follow-on missions, planets with  mass $\gsim1 M_J$  might be detected  
around double neutron stars even at cosmological
distances $z\sim 1$. In this manner, gravitational wave observation has potential to  make interesting contributions to extra-solar planetary science.

\end{abstract}

\section{Introduction}

Since gravitational wave (GW) detectors have omni-directional sensitivity,  many sources can be simultaneously observed without adjusting detectors for
individual ones. 
While this might look advantageous for astrophysical studies, it also
has downsides.  Depending on the number of GW sources, overlaps of
signals in data streams of detectors become a significant problem,
especially in the low-frequency regime probed by space GW detectors.  
For example,  LISA  (Bender et al. 1998) 
%(planned to be launched in 2015+)
 will detect
$\sim3000$ double white 
dwarf binaries  above $\sim3$mHz (see {\it e.g.} Nelemans 2006, Ruiter  et al. 2007).  Without removing the 
foreground GWs made by these numerous binaries,  it might be difficult
to observe weak interesting signals, such as extreme-mass-ratio-inspiral (EMRI) events.
In this respect,  extensive efforts are being paid to numerically
demonstrate how well both strong and weak signals can be analyzed, using mock LISA data 
(Arnaud et al. 2007).

In this paper, I study a method to search for planets (more generally sub-stellar companions) orbiting around
ultra-compact binaries. The proposed approach is to observe binaries' wobble
motions caused by the planets and imprinted as  phase modulations of
GW from the binaries. 
This approach is close to the eclipse timing method (see {\it e.g.} Deeg et al. 2008) to detect planets around binaries, and
the underlying technique 
is similar to the planet search
around pulsars with radio telescopes (Wolszczan
 \& Frail 1992, see
also Dhurandhar \& Vecchio 2001).
As the expected modulations due to planets are small,  the ongoing
numerical efforts for LISA have direct relevance to the prospects of the detection method proposed in this paper.

Here, I briefly discuss the significance of this method on
extra-solar 
planetary 
science. In the last 15 years, its rapid progress has largely been led by
theoretically unanticipated discoveries, such as those of the hot
Jupiters (Mayor \& Queloz 1995) or the pulsar planets
(Wolszczan
 \& Frail 1992).  However, at
present, observational studies 
for circum-binary planets are in a very preliminary stage (Udry et al. 2002, Muterspaugh et al. 2007, Deeg et al. 2008).
In addition, impacts of
stellar evolution processes including giant star phases or supernova
explosions are still highly uncertain (see {\it e.g.} Villaver \& Livio 2007, Silvotti et al. 2007 for recent
studies).  
Since ultra-compact binaries  such as double white dwarfs are end products of stellar evolution,  the proposed method to search for planets around them would provide us with important clues to  these
unclear issues.  While the probability of finding  a planet around a compact
binary 
is uncertain,  the large numbers of available binaries ({\it e.g.} with LISA)
 are advantageous   for various  statistical analyses,
such as
estimation of mass distribution of planets by separating information of
orbital inclination $\sin i$.

\section{Phase modulation by a planet}

To begin with, I  discuss GWs from a detached
double white dwarf binary on a circular orbit  without a planet (Takahashi \& Seto 2002). 
I write  its almost monochromatic waves around frequency $f_{gw}$
as
\beq \label{h0}%%%%%%%%%%%%%%%%%%%%%%%%%%%%%%%%%%%%%%%%%%%%%%%%%%
h_0(t)=A\cos [2\pi
f_{gw}t+\pi\dot{f}_{gw} t^2+\varphi_0+D_E(t)]\equiv A\cos[\varphi(t)],
\eeq%%%%%%%%%%%%%%%%%%%%%%%%%%%%%%%%%%%%%%%%%%%%%%%%%%%%%
where the term $(\propto {\dot f}_{gw})$ represents the intrinsic
frequency evolution  with ${\dot
f}_{gw} T_{obs}\ll f_{gw}$ ($T_{obs}$: observational time $\lsim
10$yr). The term  
$
D_E(t)\equiv {2\pi f_{gw} R_E c^{-1} \sin\theta_s}\cos[\phi(t)-\phi_s]
$
represents the Doppler phase modulations due to revolution of a detector
around the Sun
 ($R_E=$1AU) with its  orbital phase $\phi(t)=2\pi
(t/1{\rm yr})+const$.
 The angular parameters $(\theta_s,\phi_s)$ are the
direction of the binary on the sky in the ecliptic coordinate.
In eq.(1) I have neglected amplitude modulation by rotation of the
detector.   To determine the  direction of the binary, this effect is less important than the Doppler
modulation $D_E(t)$ at $f_{gw}\gsim c/R_E\sim 1$mHz
(Takahashi \& Seto 2002). { In relation to this, I do not
explicitly deal with the
orientation parameters of binaries. This is just for simplicity.  These
parameters  determine
the polarization states of the waves. }

 The 
orientation-averaged amplitude of the waves is given as 
\beqa
A&=&\frac8{\sqrt{5}}\frac{G^{5/3}\cm^{5/3}\pi^{2/3}f_{gw}^{2/3}}{r
c^4} \\
& &=6.6 \times 10^{-23}
\lmk\frac{\cm}{0.45 M_\odot}\rmk^{5/3}
\lmk\frac{f_{gw}}{3{\rm mHz}}\rmk^{2/3}
\lmk\frac{r}{8.5{\rm kpc}}\rmk^{-1}
\eeqa 
with the chirp mass
$\cm=M_1^{3/5} M_2^{3/5}(M_1+M_2)^{-1/5}$ ($M_1$ and $M_2$: two masses of
the binary).
In this equation, I put the chirp mass at ${\cal M}=0.45M_{\odot}$ (Farmer \& Phinney 2003) and used the distance to the Galactic center $r=8.5$kpc as the
typical distance to Galactic binaries. The matched filtering technique is an
advantageous method for GW observation and  the signal-to-noise ratio of the
binary is evaluated in the standard manner as
\beqa 
SNR_0&=&\frac{A\sqrt{2T_{obs}}}{h_f}\label{sn}\\
& &=138  \lmk\frac{A}{6.6 \times
10^{-23}}\rmk \lmk\frac{h_f}{1.2\times
10^{-20}{\rm {Hz^{-1/2}}}}\rmk^{-1}
\lmk\frac{T_{obs}}{10{\rm yr}}\rmk^{1/2} \nonumber 
\eeqa
with   LISA detector
noise level $h_f$   that is within 15\% around $1.2\times
10^{-20}{ \rm ~ Hz^{-1/2}}$ in the frequency regime 3mHz-10mHz relevant
for the present analysis (Bender et al. 1998). Here I assumed that LISA has two independent data streams
with  identical noise spectra.  

%From Takahashi \& Seto (2002), the typical  angular resolution (area of
%error box in the sky) of a binary is estimated as $4.0\times
%10^{-5}(SNR/138)^{-2}(f/3{\rm mHz})^{-2}$ [sr] for $T_{obs}\gsim 2$yr.

When the binary has a circum-binary planet with  mass $M_p$ and
orbital frequency 
$f_p$, the observed waveform $h_M(t)$ has an additional   phase shift $D_p(t)$
due to   the binary's wobble induced by the planet, and I put the waveform by
$h_M(t)=A\cos[\varphi(t)+D_p(t)]$. 
For a planet on a circular  orbit,  the phase shift is given by 
$
D_p(t)=\Psi_p \cos \varphi_p(t)
$  with the orbital phase 
$\varphi_p(t)=2\pi f_p t+\varphi_{c0}$  ($\varphi_{c0}$: phase constant)
and the amplitude
$\Psi_p={(2\pi  G)^{1/3} c^{-1} f_{gw} f_p^{-2/3}  M_T^{-2/3} M_p\sin
i}~$
  ($M_T=M_1+M_2$: total mass of the binary, $i$: inclination of the
planet's orbit) or explicitly
\beqa 
\Psi_p&=&0.054 \lmk\frac{M_p \sin i}{3M_J}\rmk \lmk\frac{M_T}{1.04
M_\odot }\rmk^{-2/3} \nonumber\\
& & \times  \lmk\frac{f_{gw}}{3{\rm mHz}}\rmk  \lmk\frac{f_p}{0.33{\rm
yr^{-1}}}\rmk^{ -2/3}.\label{ps}
\eeqa
For a system at a cosmological distance with redshift $z$,  the
amplitude $\Psi_p$ is given by multiplying  a factor  $(1+z)$ to eq.(\ref{ps})  with the intrinsic  (not redshifted) orbital frequency $f_p$ and the
observed 
(redshifted) GW
frequency $f_{gw}$. Note that the redshift $z$ can be estimated from
the 
observed luminosity distance (Schutz 1986).
 As I want to know the smallest
mass $M_p\sin i $ detectable with GW observation and it is easier to find a
planet with a larger amplitude $\Psi_p$, I hereafter assume $\Psi_p \ll
1$. Then the modulated signal $h_M(t)$ is expressed as 
\beq \label{hm}
h_M(t)=h_0(t)+h_{p+}(t)+h_{p-}(t)+O(\Psi_p^2)
\eeq
with two new components 
\beqa
h_{p\pm}(t)&=&{A \Psi_p} (\sin[\varphi(t)\pm \varphi_p(t)])/2 \\
&=&
-{A \Psi_p} (\sin [2\pi (f_{gw}\pm f_p)t+\pi\dot{f}_{gw}t^2+\varphi_0\pm \varphi_{p0}])/2. \nonumber
\eeqa%%%%%%%%%%%%%%%%%%%%%%%%%%%%%%%%%%%%%%%%%%%%%
 A simple interpretation can be made for eq.(\ref{hm}). In addition to
the original signal $h_0(t)$ given in eq.(1),  motion of the planet  produces
two replicas $h_{p\pm}$ (smaller by a factor of  $\Psi_p/2$ than $h_0(t)$) at nearby
frequencies $f_{gw}\pm 
f_p\gg f_p$. Because of  the coupling with the binary's rotation, the orbital
frequency $f_p$ of the planet is now 
up-converted into a band that might be observed with GW detectors. 
{ Here, it is important to note that the gravitational wave signal of
each replica $h_{p+}$ or $h_{p-}$ itself is described with a nearly monochromatic  waveform for a standard Galactic binary (including dependencies on angular parameters).  This fact is
important for  data analysis, as seen later. } 
%%%%%%%%%%%%%%%%%%%%%%%%
  In this paper I only study  a planet on a  circular
orbit, but this
analysis can be straightforwardly  extended for multiple   planets
or eccentric orbits that produce  other small replicas at frequencies $f_{gw}\pm
n f_p$ ($n=2,3,\cdots$) not only with $n=1$ (Dhurandhar \& Vecchio 2001). 
%%%%%%%%%%%%%%%%%%%%%%%

Based on the simple interpretation of the modulated signal $h_M(t)$,  I
can naively define the signal-to-noise ratio for the two small replicas
$h_{c\pm}$ by 
\beq \label{xc}
X_p=\frac{A \Psi_p}{h_f} \sqrt{T_{obs}}=5.3
\lmk\frac{\Psi_p}{0.054}\rmk  \lmk\frac{SNR_0}{138}\rmk 
\eeq
as for the original one $h_0$ given in eq.(3). If the parameters
${\vect \alpha_O}=(A,f_{gw},{\dot f }_{gw},\varphi_0,\theta_s,\phi_s)$ are
well determined  with  the
strong original one $h_0$, they can be used to estimate the three  additional 
parameters ${\vect \alpha}_N=(f_p,\Psi_p,\varphi_{c0})$ for the small
replicas $h_{p\pm}$. The 
expected observational errors for the three new parameters ${\vect \alpha}_N$ are evaluated by a $3\times
3$  Fisher matrix (see {\it e.g.} Takahashi \& Seto 2002), and I     obtain the asymptotic results at
$T_{obs}f_p\gg 1$ as
\beq \label{er}
\lmk\frac{\Delta \Psi_p}{\Psi_p}\rmk_3\equiv X_p^{-1},~~~
(\Delta f_p)_3\equiv \sqrt{3}\pi^{-1} T_{obs}^{-1} X_p^{-1},
\eeq
where the suffix ``3'' represents fitting only three new
parameters.

The actual observational  situation is more
complicated.  For example, the 
frequency resolution is given by $\sim T_{obs}^{-1}$, and with a short
observational period $T_{obs}$,  there must be a significant interference
between the  weak signals $h_{p\pm}(t)$ and  the 
strong one $h_0(t)$. { Furthermore,  the triplet
$(h_0,h_{p\pm})$ is need to be identified in the presence of thousands of other binaries.  To study the interference within the triplet,  I
firstly discuss signal analysis only with a single binary-planet system and
detector noises.}  I numerically
evaluated the observational errors expected for simultaneous fitting of
all the  nine
parameters ${\vect \alpha}_O$ and  ${\vect \alpha}_N$ listed above, and obtained the
magnitudes of the 
errors $\lmk {\Delta \Psi_p}/{\Psi_p}\rmk_S$ and $(\Delta f_p)_S$
for various sets of input parameters (suffix ``S'': simultaneous fitting).
I found that, for a given observational time $T_{obs}\gsim 1$yr,  
these results
depend strongly on the orbital frequency $f_p$, weakly on the phase 
$\phi_{c0}$, and negligibly on other parameters.  As shown in figure 1,
the errors  $\lmk {\Delta \Psi_p}/{\Psi_p}\rmk_S$ and $(\Delta f_p)_S$
become much larger than the previous simple estimations $\lmk {\Delta
\Psi_p}/{\Psi_p}\rmk_3$ and $ (\Delta f_p)_3$ for frequencies
\beq \label{sep}
f_p T_{obs}\lsim 2~~~{\rm or}~~~
|f_p-1|T_{obs}\lsim 1.
\eeq
  In the latter regime, two phase modulations
$D_E(t)$  
and $D_p(t)$ become highly degenerated.
%\footnote{We also examined the
%potential effects of amplitude modulation around frequencies
%$f_p=N~[{\rm yr^{-1}}]$ ($N(\ge 2)$: natural number). With newly including
%orientations of binaries as the fitting parameters (totally 11), we found
%that  the parameter estimation errors for the planets are almost
%unchanged around these frequencies. }.  { In the former case, the
%modulation due to the planet degenerate with the frequency evolution
%${\dot f}_{GW}$. }
%%%%%%%%%%%%%%%%%%%%%%%%%%
%If direction of the source
%$(\theta_s,\phi_s)$ is identified with electro-magnetic wave
%observation, this is no longer a problem.
%%%%%%%%%%%%%%%%%%%%5
 Outside these two bands,  the
replicas $h_{p\pm}$ are  well separated  from the original one $h_0$ in
the frequency space,  and
the simple estimation in eq.(\ref{er}) becomes reliable. 
In these preferable frequency regimes, 
 the mass $M_p \sin i$ can
be estimated  within $10\%$ error (at the same time, the naive SNR  $X_p\gsim 10$) for 
planet with $M_p\sin i\ge 5.4M_J (f/{\rm  3mHz})^{-5/3} (f_p/{\rm 0.33
yr^{-1}})^{2/3}$.  Here I used  eqs.(\ref{ps})(\ref{xc}) and (\ref{er}), and the following typical parameters:
$r=8.5$kpc, $M_1=M_2=0.52
M_\odot$, $T_{obs}=10$yr and $h_f=1.2\times
10^{-20}{\rm Hz^{-1/2}}$.

\begin{figure}[!t]
\epsscale{1.2}
\plotone{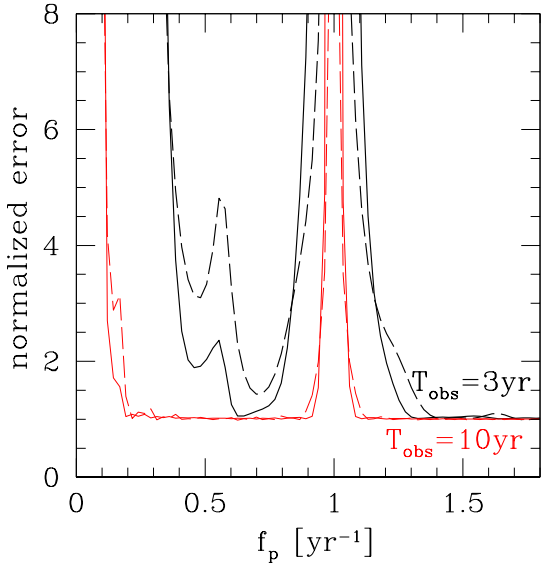}
\caption{ Planet search sensitivity around white dwarf binaries with  LISA. Estimated
 observational errors are  presented for 
 the 
 orbital 
 frequency of the 
 planet ( $(\Delta f_p)_S$: long-dashed curves) and for the amplitude of
 the  GW phase
 modulation    ($\lmk {\Delta \Psi_p}/{\Psi_p}\rmk_S$: solid curves)
 induced by 
 the planet.   These errors are normalized by their asymptotic values  $(\lmk {\Delta
\Psi_p}/{\Psi_p}\rmk_3, (\Delta f_p)_3)$
derived with   a simple interpretation for the signal
 modulation (see eq.(\ref{er})). 
 Thick curves are for integration period $T_{obs}=3$yr and thin ones for
 $T_{obs}=10$yr.  It is difficult to find a planet with a low orbital
 frequency $f_p T_{obs}\lsim2$ due to the poor frequency
 resolution. Two phase modulations $D_E(t)$ and $D_p(t)$ induced by
 motions of LISA and the planet degenerate at $|(f_p-1)|T_{obs}\lsim 1$. Outside these two bands the new signal $h_{p\pm}(t)$ by the planet can be well separated  from the strong original one $h_0(t)$, and the planet search works efficiently.  These results depend very weakly or negligibly on source parameters other than $f_p$.}
\label{fig1}
\end{figure}

Now I study  circum-binary planet searches among gravitational waves
from other binaries. For simplicity, I pick up a binary-planet system at
$f\ge 3$mHz and outside the interfering frequency regimes (\ref{sep}). I consider
 the
following two-steps data analysis; (i) detecting the individual
signals $h_0$, $h_{p+}$ and $h_{p-}$, and (ii) identifying a triplet
combination  caused by a planet.  The frequency distribution of Galactic
white dwarf binaries is modeled as $dN/df=0.08(N_B/3000)(f_{gw}/3{\rm
mHz})^{-11/3}~[{\rm yr^{-1}}]$ with the total number $N_B\sim 3000$ at
$f_{gw}\ge3$mHz (see {\it e.g.} Bender et al. 1998).  A similar density is expected  for Galactic AM CVn
stars 
(Nelemans 2006).  For  observational period $T_{obs}\sim 10$yr, the occupation number $T_{obs}^{-1}~dN/df$ of binaries per frequency bin 
will  be much  smaller than 1.   For a  planet
search, it is crucial to detect replicas $h_{p\pm}$ whose signals are
weak but individually fitted with standard Galactic binary waveforms.
Identification of weak binary signals is currently one of the most 
important topics on LISA data analysis. While the situation is somewhat
different, Crowder \& Cornish (2007) demonstrated that many (but
not all) binaries can be detected  down to $SNR\sim 7$ (corresponding to $X_p/\sqrt2\sim 7$
for each replica) even under a more crowded condition {\it i.e.} a larger
occupation number $T_{obs}^{-1}~dN/df$ (see their \S4.2 and \S4.3).   They  also showed that the Fisher matrix
analysis provides a reasonable prediction for parameter estimation errors.
 These results are very encouraging for a planet search that might   reversely provide
another motivation for ongoing activities for LISA data analysis.

Next I discuss an outline for identifying a triplet signal
by a 
binary-planet system.  The  first task is to search for a potential pair $h_0$-$h_{p+}$
from a list of 
resolved binaries, using the fact that the pair  should have same direction
(and orientation) parameters with similar frequencies.   The second task
is to confirm the existence of another replica $h_{p-}$ whose parameters
can be estimated only with the  $h_0$-$h_{p+}$ pair.  Considering the
expected binary 
density $dN/df$,  this discrimination method will  work well. In this
manner the triplet can be identified among other binaries with
a small extension of the standard Galactic binary search.  Then coherent analysis can be performed  for the modulated signal $h_m$ to improve the
quality of  parameter estimation.

For unambiguous detections of planets, other effects that produce similar waveforms should  be closely examined.
From the arguments about the triplet structure,
it is  expected  that the
phase modulation $D_p(t)$  can be easily separated from
other  small  modulations at higher frequencies $\gg f_p$ that also
generate  small replicas but with larger frequency differences.
Meanwhile, because of geometrical nature of gravitational wave generation, an observed waveform depends on angular parameters describing configuration of a binary.  For example, it is shown that, for an eccentric binary in the LISA band, an triplet waveform can be produced by the periastron advance with a frequency difference
$O\rm (1yr^{-1})$  (Seto 2001, Willems et al. 2007). But the triplet structure is different from the planet case. Precession of orbital plane of a binary (by the spin-orbit coupling) can also generate a triplet waveform, and might be important for double neutron stars with BBO/DECIGO.  But it has different amplitude patterns (or equivalently  polarization states), and has a larger frequency difference.

%\footnote{For
%example, in the case of double neutron stars for BBO/DECIGO analyzed
%later, the orbital precessions due to the spin effects have higher
%requencies. In the LISA band,  an  eccentric binary can also
%produce a  triplet waveform with a similar  caused by
%periastron advance  .  But its triplet structure is different from that analyze%d in this paper. }.  
%In addition to planets, an  eccentric binaries can 
%produce a  triplet waveform with a similar frequency difference
%$O\rm (1yr^{-1})$  caused by
%periastron advance  (Seto 2001), but they also have other orbital harmonics due% to elliptical 
%motions (Peters \& Mathews 1963).  
%Note also that a  typical close white dwarf
%binary is  expected to have a  circular orbit (Farmer \& Phinney 2003) (see als%o Willems et al. 2007).

In figure 2, I plot the detectable planet on the semimajor axis-mass
plane.
Here I used the relation   $a=1(f_p/1{\rm yr^{-1}})^{-2/3}(M_T/1M_\odot)^{1/3}$AU for the orbital frequency $f_p$ and the
semimajor 
axis $a$.  The planets around $f_p=1{\rm yr^{-1}}$ (corresponding to 1.01AU
for $M_T=1.04M_\odot$) are excluded due to the degeneracy discussed
before.

\begin{figure}[!t]
\epsscale{1.2}
\plotone{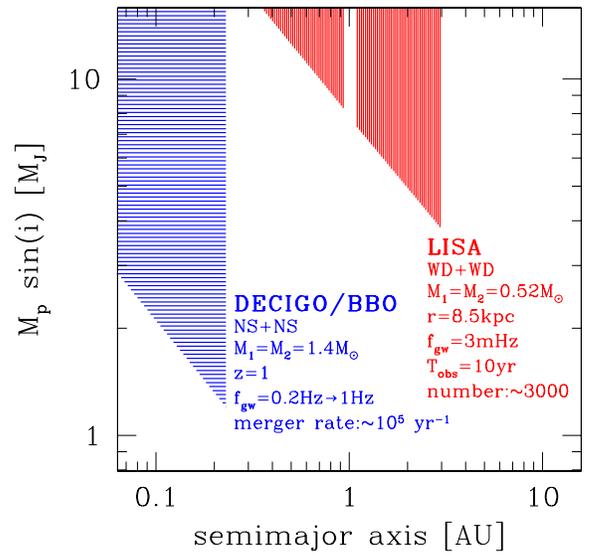}
\caption{ The ranges of detectable planets for LISA and DECIGO/BBO
 for typical sets of parameters. In the shaded regions, the mass $M_c \sin i$ can be estimated within $10\%$ error.
Around 1AU (corresponding to $f_p=1{\rm yr^{-1}}$), performance of LISA
 is degraded  due to the degeneracy of two  phase shifts induced by orbital motions of
 planet and  
 LISA.   }
\label{fig2}
\end{figure}

\section{Discussions}

The follow-on missions to LISA, such as the Big Bang Observer (BBO) (Phinney 2003) or 
the Decihertz Interferometer Gravitational Wave Observatory (DECIGO) (Seto et al. 2001, Kawamura et al. 2006)
were proposed primarily to detect stochastic GW background from
inflation in the band $f_{min}\lsim f \lsim f_{max}$ with $f_{min}\sim
0.2$Hz and $f_{max}\sim1$Hz. At the lower frequency regime $f\lsim
f_{min}$, the foreground GWs  by extra-Galactic white dwarf binaries would
fundamentally limit  sensitivity for GW observation (Farmer \& Phinney 2003).  In contrast, at $f\gsim
f_{min}$, a 
deep window of GW is expected to be opened.
To this end, it is crucial to   resolve and remove  foreground  GWs
generated by cosmological double neutron  star binaries (NS+NSs) whose estimated
merger rate is $\sim 3\times 10^5{\rm yr^{-1}}$. In addition to
NS+NSs,  there might be double
black hole 
binaries or black hole-neutron star binaries, while their merger
rates are highly uncertain.  Here I provide a brief sketch for planet search around cosmological NS+NSs  with the follow-on missions. 
I fix masses of NS+NSs at $M_1=M_2=1.4 M_\odot$.

In the observational band $[f_{min},f_{max}]$,   a NS+NS is on its final
stage 
before  merger.  The time left before the merger is
$1(f_{gw}/0.2{\rm Hz})^{-8/3}(1+z)^{-5/3}$ yr
that severely  limits the 
observable orbital frequency $f_p$ of a planet. Using the restricted 1.5-order
post-Newtonian  waveform (Cutler \& Harms 2006), I evaluated the expected observational errors
in the scenario that all the parameters are simultaneously fitted,
including two phase shifts $D_E(t)$ and $D_p(t)$.  For various sets of
input parameters, I examined the observational error for  the amplitude
$\Psi_{p,1}\equiv \Psi_p|_{f_{gw}=1{\rm Hz}}$, and found that by
observing at least three orbital cycles (namely $f_p\gsim f_{th}\equiv
3 (f_{min}/0.2)^{8/3}(1+z)^{8/3}{\rm yr^{-1}} $) the relative error is
given as
\beqa 
\lmk\frac{\Delta \Psi_{p1}}{\Psi_{p1}}\rmk &\sim&\frac{\Delta (M_p\sin i)}{M_p\sin i}\nonumber\\
&\sim& (1+z)^{-1} \lmk\frac{2.3}{SNR_0}\rmk
\lmk\frac{M_p\sin i }{3 M_J}\rmk^{-1} \lmk\frac{f_p}{3{\rm yr^{-1}}}
\rmk^{2/3} \label{er2}
\eeqa
with the signal-to-noise ratio $SNR_0$ for the observed  NS+NS. Here I assumed   a
nearly flat  noise spectrum (in units of $\rm Hz^{-1/2}$)  in the band $[f_{min},f_{max}]$ (Phinney 2003, Kawamura et al. 2006).
For a given orbital frequency 
$f_p$ and signal-to-noise ratio $SNR_0$,  the mass resolution is better than the previous results for
LISA. This is because of the higher frequencies $f_{gw}$ used in the
present case.  For $f_p \lsim f_{th}$ (less than three orbital cycles in the observational  band), the observational error ${\Delta (M_p\sin i)}/{(M_p\sin i)}$ becomes
significantly larger than eq.(\ref{er2}).

Due to a limitation of estimated computational power available  at the
time of  the
follow-on missions $\sim$2025, the minimal noise level of detectors
required to 
remove NS+NSs  
corresponds to $SNR_0\sim 100$ for  NS+NSs at $z= 1$ (Cutler \& Harms 2006).    For $z=1$ the critical orbital frequency becomes
$f_{th}=19 {\rm yr^{-1}}$ (semimajor axis $\sim 0.23$AU for $M_T=2.8M_\odot$), and the
mass $M_c \sin i$ 
 can be measured within 10\%
error for a planet  with
$M_p\sin i>1.2 (f_p/19{\rm
yr^{-1}})^{2/3}(SNR_0/100)^{-1}M_J$. The range of detectable planets is shown in
figure 2.  If detected at $z\sim 1$, the planet is $\sim 10^6$ times as
distant as those currently found in our galaxy.
 Note that the estimated merger rate of NS+NSs around $z\sim1$ is   $\sim
10^5{\rm yr^{-1}}$.  The bottom edge of the shaded region moves to (0.39AU,
0.52$M_J$) for NS+NSs at $z=0.5$.

\if0
As the lifetimes of the observed NS+NSs are very short
with BBO/DECIGO and the proposed  method based on GW
measurement  is so 
powerful, it might
not be straightforward to cross check the existence of a planet detected
first with GWs.  
If a merger of NS+NS is associated with
a  short-duration ($\ll 10$sec) burst event,  we might detect 
an  electro-magnetic 
 wave signature ({\it e.g.} reflected light) from the planet. 
If this challenging observation is achieved,  we can use the planet
as a sample 
 target to study the nature of the   bursts. 
 The signature will be delayed
by 
$\sim 100$sec from the merger time that can be predicted accurately with 
gravitational wave observation. Since we can also estimate the semimajor
axis and 
orbital phase of the planet by analyzing GWs, the measurement of the
time delay 
allows us to  geometrically
determine  the inclination angle $i$, and  then we can obtain the mass $M_p$
separately from  the factor $\sin  i$. 
\fi

I would like to thank an anonymous referee for helpful comments to improve the draft.

%\lmk\frac{}{}\rmk 

%\input{ref3}

\end{document}